# Significance of a combined approach for replacement stones in the heritage buildings' conservation frame

# Apport d'une approche combinée pour les pierres de remplacement dans le cadre de la conservation du patrimoine construit

**Olivier Rozenbaum** [a], **Luc Barbanson**[b]**, Fabrice Muller**[a] **and Ary Bruand**[a]

[a]Université d'Orléans, CNRS/INSU, université de Tours, Institut des sciences de la Terre d'Orléans (ISTO), UMR 6113, 1A, rue de la Férollerie, 45071 Orléans cedex 2, France
[b]Université d'Orléans, CNRS/INSU, université de Tours, Institut des sciences de la Terre d'Orléans (ISTO), UMR 6113, rue Saint-Amand, bât. Géosciences, BP 6759, 45067 Orléans cedex 2, France

**Abstract**

Stone substitution is a conventional operation during heritage buildings' restoration, but becomes problematic for architects and restorers when the quarry is mined out. The compatibility of the substitution stones with the original ones has been for long mainly based on the aesthetical aspect, this resulting too often in a patchwork of original and substitution stones with different patina after several years because of differences of properties. In this study, the objective is to show how substitution stones can be selected by combining aesthetic criteria and stones properties that are relevant for analyzing their compatibility. A couple of French limestones with their potential substitution stones were selected for the study. Our results showed that potential substitution stones selected on their aesthetic criteria require to be rejected because of their differences of physical properties. On the other hand, our results showed also the possibility to select substitution stones with satisfactory aesthetic aspect and properties that enable to expect a satisfactory compatibility with the original stone.

**Résumé**

Le remplacement de pierres est une opération conventionnelle lors de la restauration de bâtiments historiques, mais qui devient problématique pour les architectes et les restaurateurs quand la carrière est épuisée. La compatibilité de pierres de substitution ayant une autre origine a longtemps été appréciée sur la base de critères essentiellement esthétiques, avec trop souvent pour conséquences, après plusieurs années, une mosaïque de pierres d'origine et de substitution, avec différentes patines en raison de différences de propriétés. Dans cette étude, l'objectif est de montrer comment il est possible de sélectionner des pierres de substitution en combinant des critères esthétiques et des propriétés de ces pierres qui soient pertinents pour analyser leur compatibilité. Un couple de calcaires français, ainsi que leurs pierres de substitution potentielles, ont été sélectionnés pour l'étude. Nos résultats montrent que des pierres de substitution potentielles sur la base de critères esthétiques sont à écarter en raison de leurs propriétés physiques, qui compromettent leur compatibilité. Par ailleurs, nos résultats montrent aussi qu'il est possible de sélectionner des pierres de substitution qui présentent des critères esthétiques et des propriétés qui permettent d'atteindre une compatibilité satisfaisante avec les pierres d'origine.

## 1. Introduction

Heritage buildings are very sensitive to their environment that may alter them to the point of their whole destruction. Environmental parameters such as the frequency and intensity of rains and freezing/thawing cycles, as well as the nature and concentration of pollutants in rains and atmosphere influence the alteration processes [8] and [25]. Techniques were developed to restore the weathered stones of heritage buildings: water washing and abrasive cleaning by powder projection [9] and [28], laser cleaning [5], or reinforcement by chemical products (resin, water-repellent treatment) [1] and [26]. When the stones are deeply weathered, they require to be replaced by new ones [11], [13], [14], [15], [16] and [27]. Usually, the restorers select substitution stones that originate from the quarry from where the weathered stones were originally extracted. Databases relating stones of heritage buildings to the quarries from where they were extracted [2], [4], [12] and [13] or locating the types of stones still available in quarries [4], [12], [14] and [17] were established. When the quarry is no longer accessible or mined out, or when the origin of the stones remains unknown, the restorers select substitution stones originating from other quarries in similar geological formations. The compatibility of the substitutions stones with the original ones has been for long mainly based on the aesthetical aspect (colour, grain size, fossils, texture). Recently, Dreesen et al. [14] and Elert et al. [15] pointed out the necessity to study also the composition and microstructure of the substitution stones. Dessandier et al. [11] compared eighteen building stones and discussed their physical properties. Dreesen et al. [14] and Hyslop et al. [20] reported that inadequate replacement can result in a patchwork of original and substitution stones weathered in several ways after several years because of differences of physical properties. Finally, Henriques [19] and Sasse and Snethlage [24] discussed stone compatibility when restoring heritage buildings and particularly the main physical properties that are to be taken into account. Henriques [19] and Sasse and Snethlage [24] showed that the main physical properties were the hydraulic mechanical properties. The objective of our study is to show how substitution stones can be selected by combining aesthetic criteria, stones characteristics and properties that are relevant for analyzing their compatibility. A couple of French limestones, largely used as building stones in France, with their potential substitution stones, were selected for the study.

## 2. Material and methods

One of the French limestones selected for this study was the Saint-Maximin stone (Lutetian) [10] that originates from the North of the Paris Basin and was largely used in order to build monuments in a radius of about 200 km around Paris. Amongst the numerous constructions, one can mention cathedrals (Le Mans, Meaux), churches (Caen, Rouen), castles (Saint-Cloud, Vincennes) and many buildings and monuments, as well as several railway stations in Paris. Although the Saint-Maximin stone is still available, a possible Romanian substitution limestone (stone A) originating from the Transylvanian Basin was selected for our study on the basis of aesthetical criteria (similar textural aspect with coarse grains visible to the naked eye). On the other hand, the second French limestone selected was the Garchy stone that originates from the South of the Paris basin (Oxfordian) [22]. This stone was largely used for many constructions in a radius of about 200 km around the quarries and can be found in cathedrals (Orleans, Nevers, Bourges), castles (Nozet in Pouilly-sur-Loire, Verger in Suilly-la-Tour, Sancerre), the Beaugency bridge, the Orleans art school, or the Caen chamber of commerce. As for the Saint-Maximin stone, the Garchy stone is still available, but a possible substitution limestone (stone B) originating from central Romania was also selected on the basis of aesthetical criteria for the study (similar colour and textural aspect with fine grains visible to the naked eye).

The averaged chemical composition and mineralogical composition were determined by using respectively induced coupled plasma (ICP), infrared spectroscopy and powder X-ray diffraction. The clay mineralogy was specifically studied by using powder X-ray diffraction. For this latter study, diffraction patterns (not presented in here) were recorded on the finest stone particles. This was achieved by grinding fragments manually in a mortar with a pestle in order to form a < 200-µm powder. Then, $CaCO_3$ was removed adding drop by drop 1 N HCl to a suspension of the < 200-µm powder in water. The resulting material was washed to eliminate soluble salts by using successive cycles of decantation by centrifugation and dispersion in water. The < 2-µm was then collected on the basis of the Stokes law and oriented on a glass slice before analysis by X-ray diffraction.

Thin sections parallel to the sediment bedding were obtained after resin impregnation. They were polished and observed in optical microscopy and then coated with carbon prior to observation in scanning electron microscopy by using the backscattered electron emission [7]. Thus, the recorded images enabled the discussion of the averaged grain and pore size and of the microscopic stone texture [23]. Fresh surfaces of broken stones were also observed by using the secondary electron emission in scanning electron microscopy. Furthermore, analysis of the porosity in microscopy was complemented by mercury intrusion porosimetry. The latter was performed with a porosimeter that operated from a pressure of 4 kPa up to a maximum of $2 \times 10^3$ kPa, enabling pore size distribution study for pores with apparent diameters ($D_a$) ranging from 360 down to 0.006 µm, respectively. The mercury intrusion curve was analysed according to Bruand et al. [6] and Bruand and Prost [7]. Thus every pore class was identified by its modal apparent diameter ($D_{a,m}$). Values for the surface tension of mercury and its contact angle on solid material were 0.484 N m$^{-1}$ and 130°, respectively. Finally, specific surface area measurement was carried out by physical adsorption of nitrogen gas molecules (BET).

Transfer properties were described by determining the stone imbibition coefficients [3], [18] and [21]. The lower surface of a stone was placed in contact with water and, due to the preferential wetting of the solid, the water fills the pores, pushing the air inside the pores out of the sample. The height of the capillary front and the water mass uptake were then measured as a function of time. Assuming the pores cylindrical and neglecting the gravity effect on water, the mass uptake ($\Delta m$) per surface area unit ($S$) and the capillary height ($h$) are function of the square root of time:

$$\Delta m/S = A\sqrt{t},$$

$$h = B\sqrt{t},$$

where the imbibition coefficients $A$ and $B$ are defined as following:

$$A = \pi r^2 \sqrt{\frac{r\gamma\cos\alpha}{2\eta}}$$

$$B = \sqrt{\frac{r\gamma\cos\alpha}{2\eta}}$$

with $r$ the radius of the capillary, $\eta$ the water viscosity, $\gamma$ the superficial tension and $\alpha$ the angle between the solid and the fluid. An effective hydraulic lattice could be constituted if the porous media is considered as formed by $N$ vertical capillaries with a radius $r_{eq}$. This latter was derived from Eq. (4):

$$r_{eq} = \frac{2\eta B^2}{\gamma\cos\alpha}$$

and $N$ was given by [21]:

$$N = \frac{A}{\pi B r_{eq}^2}$$

Measurements were performed on cylindrical samples (diameter: 50 mm), but small enough to neglect the gravity effect (height: 100 mm) [21]. The samples were oven-dried for 96 h at 50 °C and put in a desiccator with phosphorous anhydrite in order to reach room temperature while maintaining a dry environment. In order to take into account the anisotropic characteristics of the stones, the cylinders were cut parallel and perpendicular to the sediment bedding. Capillary ascent and mass taking are represented versus the square root of time. Thus, according to Eqs. (1) and (2), it was expected for homogeneous stones (*A* and *B* constant) to record an increase in Δ*m* and *h* with respect to √t[3], [18] and [21]. Furthermore, an anisotropy coefficient (($1-P\perp/P_{//}$)×100; *P*=*A* or *B*) was computed in order to discuss the differences of properties between the two orientations with respect to the sediment bedding.

Uniaxial compressive strength tests were also performed in order to determine the maximum value of stress reached before failure. Cylindrical samples 100 mm high and 50 mm in diameter were cut parallel and perpendicular to the sediment bedding in order to obtain information about mechanical isotropy even if, nowadays, buildings stones are rarely placed perpendicularly to the sediment bedding. An anisotropy coefficient was also computed for the mechanical resistance, as done above for the imbibition properties. Cylindrical samples were oven-dried and put in a desiccator as for the imbibition measurements. Then, the measurements were performed at a constant loading speed rate (0.3 mm min$^{-1}$), and if *F* is the maximal force just before the failure, the uniaxial compressive strength is given by the following formula:

$$R_c = \frac{F}{\pi(d/2)^2}$$

where *d* is the cylinder diameter. For imbibition and compressive strength measurements, five samples were selected for each stone and experiment. This enabled to record information about stone heterogeneity.

## 3. Results and discussion

### 3.1. Morphological, chemical and textural analysis

Macroscopically, the Saint-Maximin stone and stone A were quite similar (aspect, size and repartition of the grains), both being largely heterogeneous with coarse grains. The Saint-Maximin stone and stone A were yellowish-white coloured, the Saint-Maximin stone being slightly more ochre coloured and showing macroscopic stratigraphy. Powder X-ray diffraction and infrared spectroscopy showed that these stones were made up of calcite essentially with some silica in the Saint-Maximin stone. Chemical analysis confirmed these qualitative measurements (Table 1). X-ray diffraction of the <2 μm material after $CaCO_3$ removal showed the presence of very few clays in the Saint-Maximin stone in comparison with stone A. Clay minerals were smectites, recognized with a peak at 1.5 nm, shifting down to 0.97 nm after heating at 150 °C because of dehydration of the clay layer.

Microscopy (Fig. 1 and Fig. 2) showed the presence of numerous fossils that are more numerous and larger in the Saint-Maximin stone (foraminifera, rare algae, and spicules) than in stone A (foraminifera, algae, and bryozoa). The Saint-Maximin stone was composed of calcite grains 50 to 200 μm in size, homogeneously distributed and bonded together by micritic and microsparitic calcite (Fig. 1). The average calcite grain size in stone A (200 to 500 μm) was bigger than in the Saint-Maximin stone (50 to 200 μm), some grains being larger than 1 mm in the former. These grains were bonded together by microsparitic calcite (Fig. 2). Quartz grains were also present within the Saint-Maximin (not present in the stone A) and were embedded by calcite. The backscattered and secondary emission images of the Saint-Maximin stone showed numerous macropores 50 to 700 μm in size. Macropores were also present in stone A, but they were smaller and less numerous than in the Saint-Maximin stone, probably because of the closer grain assemblage and better cementation by calcite in stone A. Such a difference between the two stones was consistent with the data from mercury intrusion porosimetry measurements (Fig. 3 and Table 1). Indeed, the total porosity was 38% (total pore volume of 230 mm$^3$ g$^{-1}$) and 27% (total pore volume of 139 mm$^3$ g$^{-1}$) for the Saint-Maximin stone and stone A, respectively. Analysis of the cumulative pore volume curve showed that most porosity of the Saint-Maximin stone corresponded to pores with $D_{a,m}$ = 18 μm (181 mm$^3$ g$^{-1}$) and secondarily to smaller pores with $0.1 \leq D_{a,m} \leq 1$ μm (49 mm$^3$ g$^{-1}$). On the other hand, most porosity of stone A corresponded to pores with $D_{a,m}$ = 8 μm (73 mm$^3$ g$^{-1}$) and secondarily to pores with $0.1 \leq D_{a,m} \leq 1$ μm (52 mm$^3$ g$^{-1}$), thus showing similar pore volumes for the small pores (i.e. those with $0.1 \leq D_{a,m} \leq 1$ μm) of the Saint-Maximin stone and stone A. For both stones, specific surface areas were close and rather weak (Table 1).

At the macroscopic scale, the Garchy stone and stone B were also quite similar. They were yellowish-grey veined, with finer grains than in Saint-Maximin stone and in stone A. Powder X-ray diffraction and infrared spectroscopy showed similar composition, with $CaCO_3$ essentially. Infrared spectroscopy showed very small amounts of $SiO_2$ in stone B. Chemical analysis confirmed these qualitative measurements (Table 1). The <2 μm material of the Garchy stone showed the presence of a peak at 0.72 nm, indicating the presence of kaolinite and of a large peak between 1 and 2 nm, indicating the presence of interstratified 2:1 clay minerals. At 150 °C, this peak was shifted down to 0.99 nm because of clay layer dehydration. After heating at higher temperature (550 °C, 2 h), the characteristic kaolin peak disappeared, confirming kaolinite presence. Stone B did not show any peak at 0.72 nm, but peaks at 1.5 and 1.0 nm, indicating the presence of smectite and micas or illite, respectively. At 150 °C, these peaks were shifted down to 0.99 nm because of clay layer dehydration.

Optical microscopy and backscattered emission images (BESI) (Fig. 4 and Fig. 5) showed that the Garchy stone was essentially composed of rounded-shaped ooliths 0.1 to 1 mm in diameter and of a few fossils (foraminifera and rare spicules). The ooliths were cemented by sparitic calcite, some of them being large (between 0.5 and 1 mm). Stone B presented numerous fossils (foraminifera) 0.5 to 1 mm in size, fulfilled by calcite and cemented by micritic calcite. There were approximately as much fossils in stone B as ooliths in the Garchy stone. Furthermore, the cemented matrix of these two stones was quite different, stone B having smaller grains than the Garchy one. Quartz grains were in very small proportion in stone B when they were absent in the Garchy stone. BESI showed that the Garchy stone was more macroporous than stone B, the former having larger pores than the latter. However, the two stones had similar total porosity with 19.7 % (total pore volume of 91 mm$^3$ g$^{-1}$) and 18.2 % (total pore volume of 82 mm$^3$ g$^{-1}$) for the Garchy stone and stone A, respectively (Table 1). The difference of pore size observed on BESI is consistent with the difference of $D_{a,m}$ recorded for the two stones: $D_{a,m}$ = 1.3 μm for the Garchy stone and 1 μm for stone B (Fig. 3). For both stones, specific surface areas were similar and quite weak (Table 1).

## 3.2. Water transfer properties

In spite of the cylindrical capillary hypothesis, the imbibition curves (Fig. 6) recorded for the samples of the Saint-Maximin stone and stone A cut parallel and perpendicularly to the sediment bedding showed a quasi linear behaviour within zone 1 (between 0 and $t_1$), thus indicating an homogeneous porous lattice [3], [18] and [21]. Within zone 2, i.e. after the visual saturation at $t_1$, when the capillary front reaches the top of the sample, the mass uptake continues and stabilizes at $t_2$. This time lag ($\Delta t = t_2 - t_1$) shows that the stones continue to absorb water by other capillary pores. This might be related to the pore size distribution characteristics and particularly to the connectivity between the pores. Within zone 3, after $t_2$, the mass uptake continues very slowly and corresponds to the infilling of the trapped porosity thanks to air diffusion through water. This diffusion process is controlled by Fick's law. The time lag was very small for the Saint-Maximin stone ($\Delta t_{//} = t_{2//} - t_{1//} \approx 2$ min and $\Delta t_{\perp} = t_2 - t_{1\perp} \approx 13$ min for tests lasting for about 70 min) and greater for stone A ($\Delta t_{//} \approx 33$ min and $\Delta t_{\perp} \approx 50$ min for tests lasting for about 300 min). The imbibition coefficients confirmed that the imbibition kinetics of the Saint-Maximin stone was far much greater than those recorded for stone A (Table 1). This result was consistent with the SEM observations and mercury porosimetry data that showed larger pores for the Saint-Maximin stone than in stone A (the imbibition coefficients were proportional to $r^{\alpha}$, $\alpha > 0$). Furthermore, for the two stones, $t_{1\perp}$ (resp. $t_{2\perp}$) was always greater than $t_{1//}$ (resp. $t_{2//}$), showing that imbibitions were always quicker through the sediment bedding direction, whatever the stone. This was consistent with the imbibition coefficient anisotropies determined between the stones cut parallel and perpendicular to the sediment bedding. Thus, the porous lattice was better connected in the sedimentation plane than perpendicularly to this plan. Moreover, it should be noted that this anisotropy with respect to the hydric behaviour was significant for the two stones studied. The radius $r_{eq}$ of an effective lattice (Eq. (5)) and the number $N$ of vertical capillaries (Eq. (6)) were reported in Table 1. Obviously, the more $D_{a,m}$ was, the more $r_{eq}$ was. Nevertheless, these two radiuses were not corresponding. This was mainly related to the complexity of the porous media (tortuosity, series of pores, pore throats, narrow parts…) that was far from being cylindrical capillaries alone. Actually, $r_{eq}$ should be only taken as a hydraulic parameter and not as a convenient estimator of pore size.

The imbibition curves (Fig. 7) recorded for the Garchy stone and stone B showed a quasi-linear increase in the first zone whatever the sample orientation with respect to the sediment bedding, a time lag defining a second zone and a third zone where the mass uptake was quasi-equal to zero, as for the previous samples. However, the imbibition coefficients of the Garchy stone and stone B were similar for the samples cut parallel and perpendicular to the sediment bedding (Table 1), showing a very slight anisotropy with respect to the hydric behaviour. For the Garchy stone, $t_{1//} \approx t_{1\perp}$ and $t_{2//} \approx t_{2\perp}$. This led to similar time lags for the two orientations ($t_{//} \approx t_{\perp} \approx 140$ min for tests lasting for about 1500 min). This was different for stone B ($t_{1//} \neq t_{1\perp}$ and $t_{2//} \neq t_{2\perp}$), even if the time lags were similar according to the orientations ($\Delta t_{//} \approx 370$ min and $\Delta t_{\perp} \approx 350$ min for tests lasting for about 2000 min), thus showing that this stone was far much longer to saturate than the Garchy one. This difference of time lag between the two stones would be mainly related to the difference of $D_{a,m}$ between them, the latter being three times as much as its value for stone B. Thus, this results in a longer time for the stone B to reach zone 3. However, the imbibition coefficients were close and the linearity of the curves indicated that the porous lattices were homogeneous on the whole sample for the two stones. Concerning the analysis by an effective hydraulic lattice (Eqs. (5) and (6)), the conclusions were similar to the previous two stones, but showing this time rather analogous porous characteristics and water transfer properties.

### 3.3. Mechanical properties

For the Saint-Maximin stone and stone A, the mechanical resistance recorded was low (Table 1). This is related to the weak cementation of the grains by the micritic and microsparitic calcite. Results also showed that the mechanical resistance was greater perpendicularly to the sediment bedding than to the parallel one. The compressive strength of stone B was largely greater than for the Garchy stone, and there was no difference according to the sediment bedding for every stone (Table 1). This difference of compressive strength reflected differences of cementation degree, as shown in microscopy.

## 4. Conclusion

Our results showed that it is possible to reveal differences of compatibility between potential substitution stones that were selected mainly on aesthetical characteristics in comparison with the stones to be substituted. The Saint-Maximin stone and stone A showed a potential incompatibility because of difference of transfer properties, as revealed by the imbibition measurements, and secondarily differences of chemical composition, porosity and pore size distribution. Thus, the utilization of stone A as a possible substitution stone for the Saint-Maximin one should be avoided.

Our results showed also that stone B might be used as a substitution stone for the Garchy stone, since the two stones present close composition and properties. The imbibition properties are indeed comparable with slight anisotropy, even if the pore size distribution showed some differences. A smaller modal apparent diameter for the pores of stone B would explain the slower water absorption recorded.

Thus, our results showed that potential substitution stones, such as stone A for the Saint-Maximin stone, should be rejected on the basis of their physical properties. On the other hand, substitution of the Garchy stone by stone B is possible, but will require to be confirmed by further experiments in climatic chambers or in situ validation. This will enable us to analyse the possible consequences of the differences in uniaxial compressive strength and water absorption properties that were recorded.

# Figures and Tables

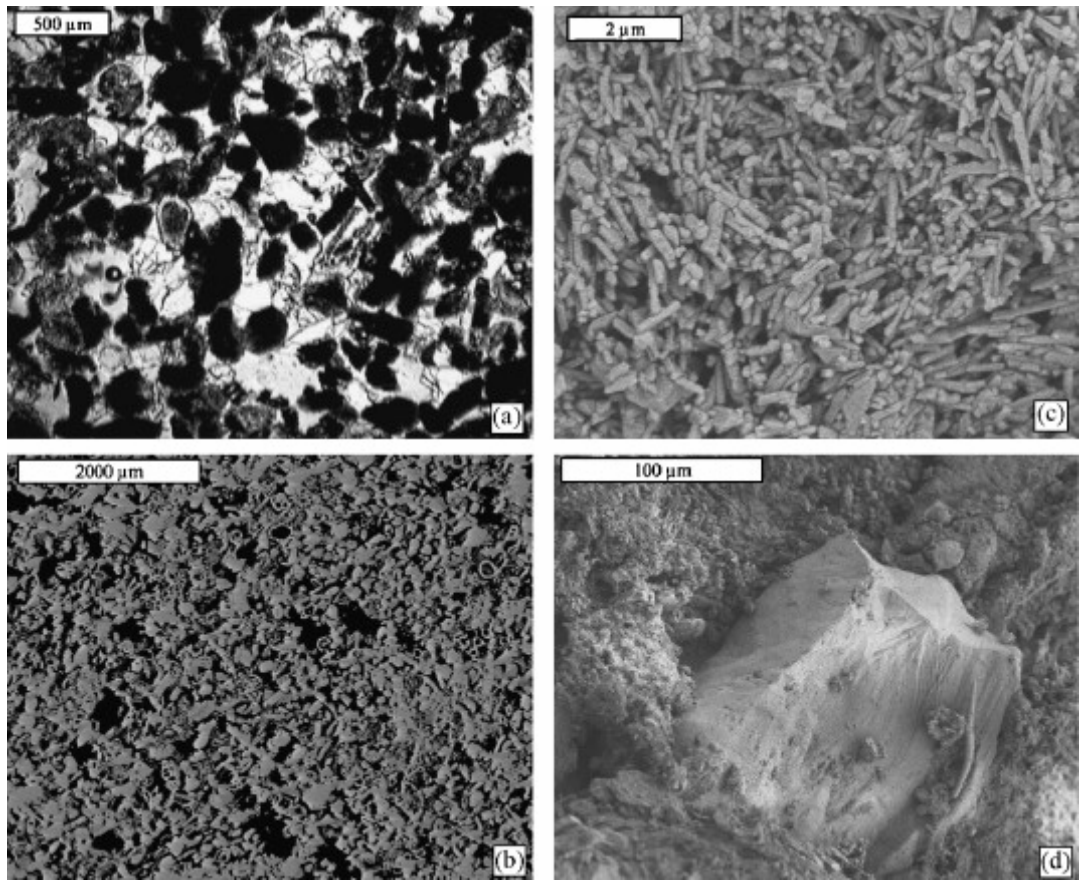

Fig. 1. Saint-Maximin stone: thin section observed in optical microscopy (**a**, transmitted polarized non-analyzed light) and scanning electron microscopy (**b**, backscattered electron); and broken stone surface observed in scanning electron microscopy (**c** and **d**), secondary electron). Voids are white to light grey in (**a**) and black to dark grey in (**b**).Fig. 1. Pierre de Saint-Maximin : lame mince observée par microscopie optique (**a**, lumière transmise polarisée non analysée) et microscopie électronique à balayage (**b**, électrons rétrodiffusés) et fracture observée par microscopie électronique à balayage (**c** et **d**, électrons secondaires). Les pores vont du blanc au gris clair dans (**a**) et du noir jusqu'au gris foncé dans (**b**).

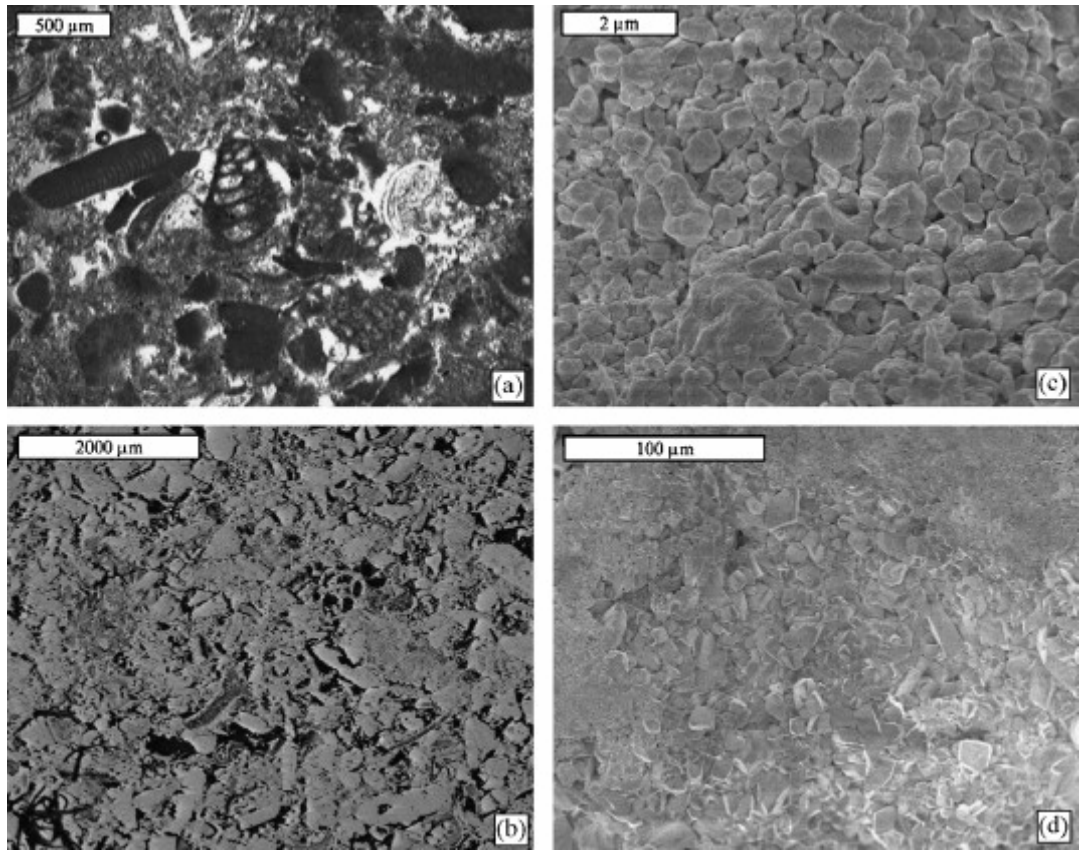

Fig. 2. Stone A: thin section observed in optical microscopy (**a**, transmitted polarized non-analyzed light) and scanning electron microscopy (**b**, backscattered electron); and broken stone surface observed in scanning electron microscopy (**c** and **d**, secondary electron). Voids are white to light grey in (a) and black to dark grey in (b).Fig. 2. Pierre A : lame mince observée par microscopie optique (a, lumière transmise polarisée non analysée) et microscopie électronique à balayage (**b**, électrons rétrodiffusés et fracture observée par microscopie électronique à balayage (**c** et **d**, électrons secondaires). Les pores vont du blanc au gris clair dans (**a**) et du noir jusqu'au gris foncé dans (**b**).

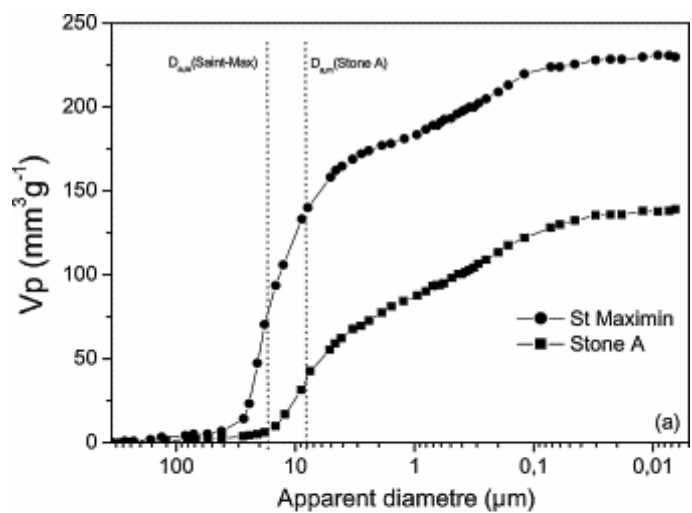

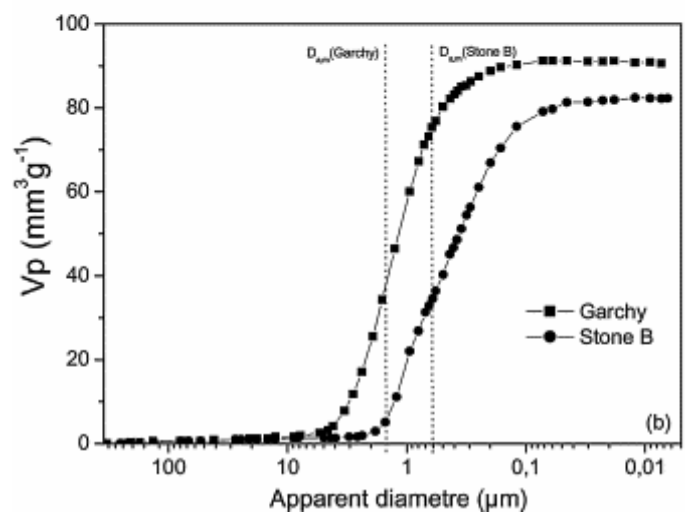

Fig. 3. Cumulative pore volume curve from mercury porosimetry.Fig. 3. Courbes de porosimétrie à mercure (volume cumulé).

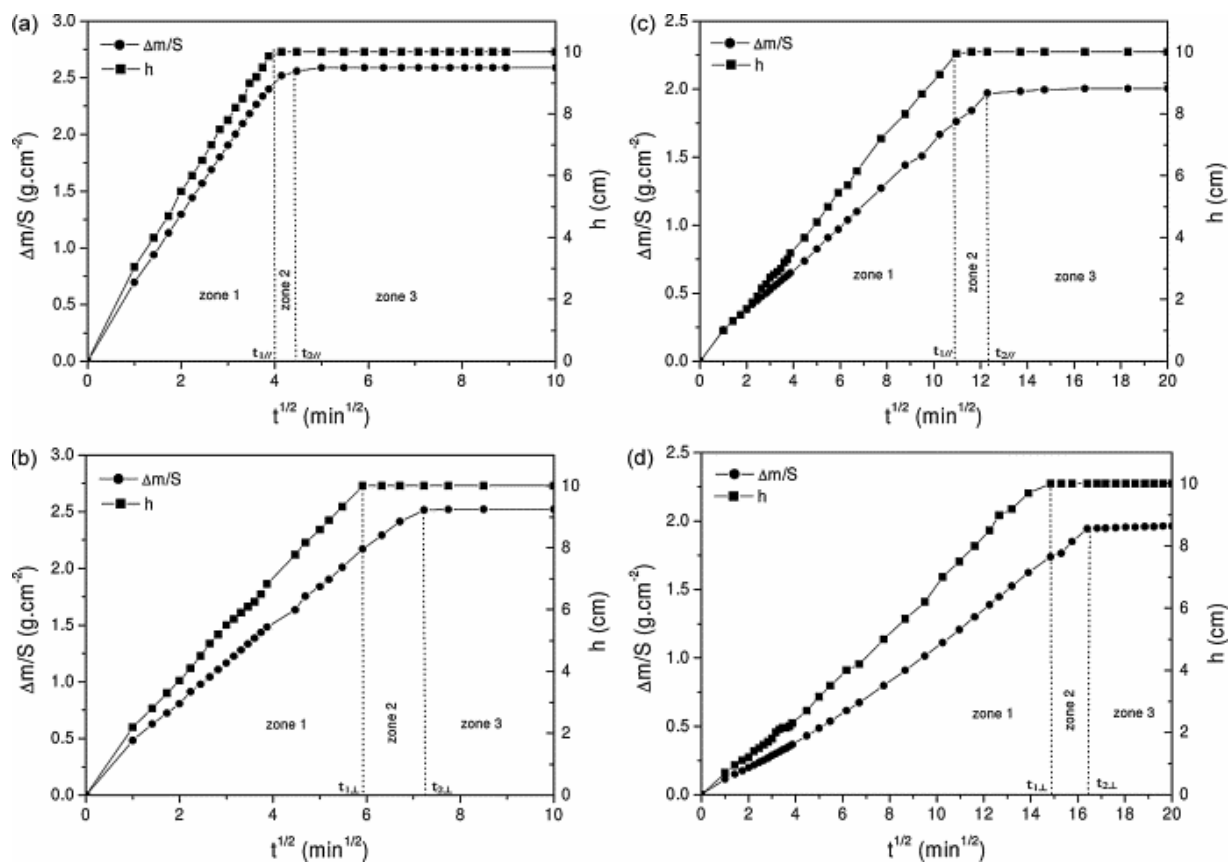

Fig. 4. Imbibition curves of the Saint-Maximin for samples taken parallel (**a**) and perpendicularly (**b**) to the stone bedding and stone A for samples taken parallel (**c**) and perpendicularly (**d**) to the stone bedding.Fig. 4. Courbes d'imbition de la pierre de Saint-Maximin pour des échantillons pris parallèlement (**a**) et perpendiculairement (**b**) au lit de la pierre, et de la pierre A pour des échantillons pris parallèlement (**c**) et perpendiculairement (**d**) au lit de la pierre.

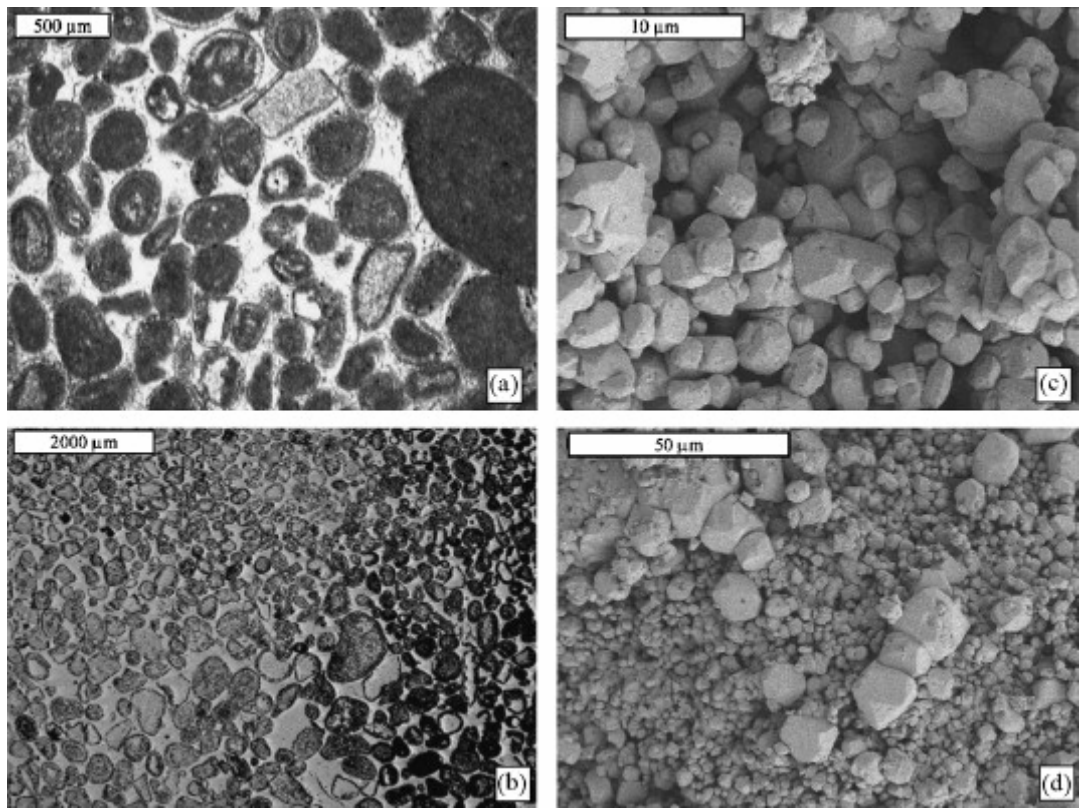

Fig. 5. Stone B: thin section observed in optical microscopy (**a**, transmitted polarized non-analyzed light) and scanning electron microscopy (**b**, backscattered electron); and broken stone surface observed in scanning electron microscopy (**c** and **d**, secondary electron). Voids are white to light grey in (**a**) and black to dark grey in (**b**).Fig. 5. Pierre B : lame mince observée par microscopie optique (**a**, lumière transmise polarisée non analysée) et microcopie électronique à balayage (**b**, électrons rétrodiffusés) et fracture observée par microcopie électronique à balayage (**c** et **d**, électrons secondaires). Les pores vont du blanc au gris clair dans (**a**) et du noir jusqu'au gris foncé dans (**b**).

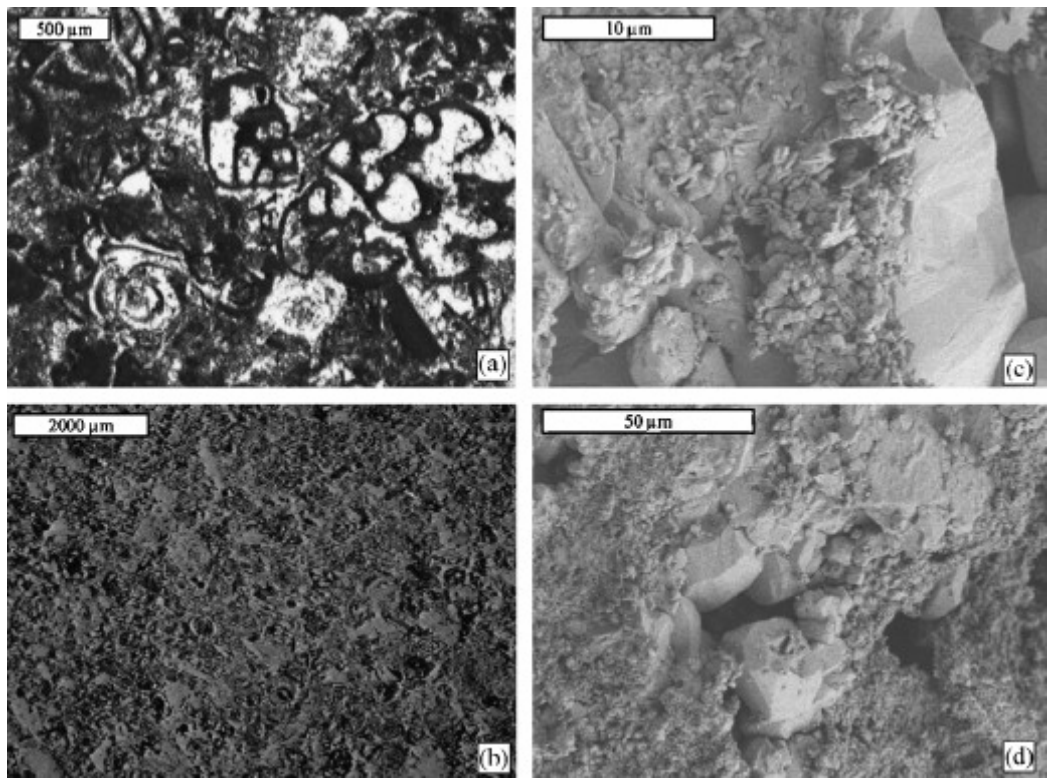

Fig. 6. Garchy stone: thin section observed in optical microscopy (**a**, transmitted polarized non-analyzed light) and scanning electron microscopy (**b**, backscattered electron); and broken stone surface observed in scanning electron microscopy (**c** and **d**, secondary electron). Voids are white to light grey in (**a**) and black to dark grey in (**b**).Fig. 6. Pierre de Garchy : lame mince observée par microscopie optique (**a**, lumière transmise polarisée non analysée) et microscopie électronique à balayage (**b**, électrons rétrodiffusés); et fracture observée par microscopie électronique à balayage (**c** and **d**, électrons secondaires). Les pores vont du blanc au gris clair dans (**a**) et du noir jusqu'au gris foncé dans (**b**).

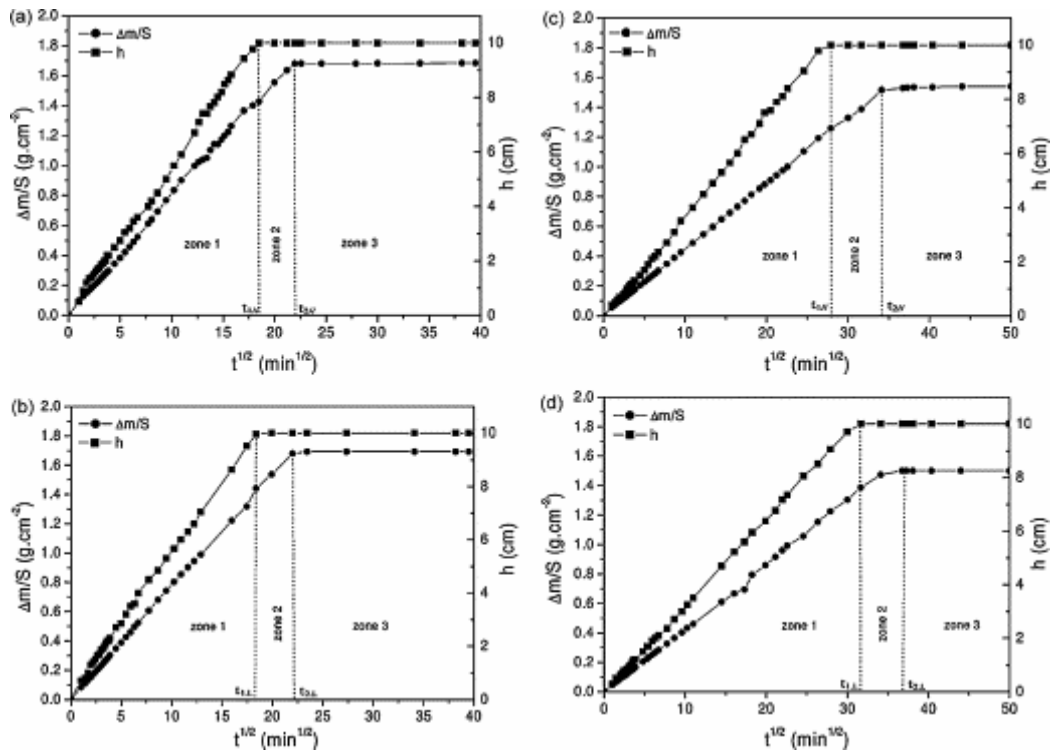

Fig. 7. Imbibition curves of the Garchy limestone for samples taken parallel (**a**) and perpendicularly (**b**) to the stone bedding and stone B for samples taken parallel (**c**) and perpendicularly (**d**) to the stone bedding.Fig. 7. Courbes d'imbibition de la pierre de Garchy pour des échantillons pris parallèlement (**a**) et perpendiculairement (**b**) au lit de la pierre, et de la pierre B pour des échantillons pris parallèlement (**c**) et perpendiculairement (**d**) au lit de la pierre.

Table 1. : Main characteristics of the four stones studied

Tableau 1 Principales caractéristiques des quatre pierres étudiées

| | Saint-Maximin | | Stone A | | Garchy | | Stone B | |
|---|---|---|---|---|---|---|---|---|
| Location | France | | Romania | | France | | Romania | |
| Chemical analysis | | | | | | | | |
| $CaCO_3$ (%) | 87.4 | | 97.1 | | 99.6 | | 97.5 | |
| $SiO_2$ (%) | 12.5 | | 0 | | 0 | | 1.7 | |
| Other (%) | 0.1 | | 2.9 | | 0.4 | | 0.8 | |
| Total porosity (%) | 38.0 | | 27.0 | | 19.7 | | 18.2 | |
| Specific surface ($m^2\ g^{-1}$) | 1.58 | | 1.75 | | 0.43 | | 0.52 | |
| Imbibition coefficient | // | | // | | // | | // | |
| $A$ (g cm$^{-2}$ min$^{-1/2}$) | 0.62 ± 0.04 | 0.34 ± 0.02 | 0.15 ± 0.01 | 0.12 ± 0.01 | 0.08 ± 0.01 | 0.08 ± 0.01 | 0.04 ± 0.01 | 0.04 ± 0.01 |
| $B$ (cm min$^{-1/2}$) | 2.51 ± 0.18 | 1.73 ± 0.09 | 0.92 ± 0.07 | 0.69 ± 0.07 | 0.54 ± 0.04 | 0.52 ± 0.05 | 0.37 ± 0.03 | 0.31 ± 0.03 |
| $\Delta t$ (min) | 2 | 13 | 33 | 50 | 140 | 140 | 370 | 350 |
| $r_{image}$ (μm) | 50 to 700 | | 10 to 300 | | 10 to 500 | | 5 to 200 | |
| $D_{a,m}$ (μm) | 18 | | 8 | | 1.3 | | 1 | |
| $r_{eq}$ (μm) | 0.288 | 0.137 | 0.039 | 0.022 | 0.013 | 0.012 | 0.006 | 0.004 |
| $N \times 10^6$ | 0.09 | 0.33 | 3.44 | 11.62 | 26.39 | 31.87 | 87.38 | 211.66 |
| $D_{a,m}/r_{eq}$ | 38.1 | 80.1 | 206.1 | 366.4 | 112.2 | 120.9 | 95.6 | 136.1 |
| | | | | | | | | |
| Anisotropy | | | | | | | | |
| $A$ (%) | 45.2 | | 20 | | 0 | | 0 | |
| B (% | 31.1 | | 25 | | 3.9 | | 16.2 | |

|  | Saint-Maximin | | Stone A | | Garchy | | Stone B | |
|---|---|---|---|---|---|---|---|---|
| Compressive strength $R_c$ (MPa) | // | | // | | // | | // | |
| | 7.4 ± 0.5 | 10.9 ± 0.9 | 12.4 ± 0.8 | 16.9 ± 1.2 | 21.2 ± 1.4 | 22.2 ± 1.4 | 39.8 ± 4.0 | 40.4 ± 4.3 |
| Anisotropy (%) | 47.3 | | 36.3 | | 4.5 | | 1.5 | |